\documentstyle[aps,prl,multicol,epsf]{revtex}
\def \BE {\begin{equation}}
\def \EE {\end{equation}}
\def \BEA {\begin{eqnarray}}
\def \EEA {\end{eqnarray}}

\def \one {^{(1)}}
\def \two {^{(2)}}

\def \e {{\epsilon}}

\begin{document}
\title{Anomalous probability of large amplitudes in wave turbulence}
\author{Yeontaek Choi$^*$, Yuri V. Lvov$^\dagger$, Sergey Nazarenko$^*$
and Boris Pokorni$^\dagger$}
\address
{
$^*$
Mathematics Institute, The University of Warwick,  Coventry, CV4-7AL, UK \\
$^\dagger$
Department of Mathematical Sciences, Rensselaer Polytechnic Institute,
Troy, NY 12180 }

\maketitle

\begin{abstract}
{Time evolution equation for the Probability Distribution Function
(PDF) is derived for system of weakly interacting waves. It is shown
that a steady state for such system may correspond to strong
intermittency.}
\end{abstract}

\begin{multicols}{2}

{\bf Introduction ---}
Wave Turbulence (WT) is a common name for the fields of dispersive
waves which are engaged in stochastic weakly nonlinear interactions
over a wide range of scales.  Numerous examples of WT are found in
oceans, atmospheres, plasmas and Bose-Einstein
condensates~\cite{ZLF,Ben,GS,Zakfil,hasselman,zaslav,OsbornePRL,JaLubimii,DNPZ}.
For a long time, describing and predicting the energy spectra was the
only concern in WT theory. More recently, some attention was given to
the study of turbulence intermittency.  WT intermittency, or
``burstiness'' of the turbulent signal, was observed experimentally
and numerically and was attributed, as in most turbulent systems, to
the presence of coherent structures. Examples include collapsing
filaments in Bose-Einstein condensates with attractive potentials
\cite{DNPZ,Bradley}, condensate quasi-solitons in systems with
repulsive potentials \cite{DNPZ,Pitaevsky,zgpd}, white caps of sea
waves at small scales \cite{rough}, freak ocean waves at larger scales
\cite{jansen}.  Often, such coherent structures are intense but quite
sparse so that in most of the space waves remain weakly nonlinear and
mostly unaffected by these structures.

Recent analysis of the higher order cumulants \cite{biven} showed that
WT becomes strongly non-Gaussian at the same length scale where it
fails to be weakly nonlinear. In scale invariant systems, the ratio of
nonlinear time to the linear wave period grows as a power-law either
in to small or toward large wavenumbers. When this growth coincides
with the cascade direction then one expects the WT breakdown if the
inertial range is large enough. Otherwise intermittency never occurs
provided that turbulence is weak at the forcing scale \cite{cnn}.
Further, even if a significant non-Gaussianity occurs, it does not in
itself imply intermittency because PDF may remain, in principle, of
the same order as Gaussian in all of its parts. This motivates us to
study PDFs in WT. The first study of PDF in the WT context was done in
\cite{zaslav} for the 3-wave systems, whereas here we will be
concerned with the 4-wave case.  We are also motivated by a puzzling
numerical evidence of a low-wavenumber intermittency in the system of
water-surface gravity waves \cite{yoko} whereas the analysis of
\cite{biven} predicts intermittency at high wavenumbers only.
Explaining this fact could shed light on the phenomenon of freak waves
\cite{jansen}.

The idea of the present letter is based on the observation that even
if the ``hard'' breakdown (as in \cite{biven}) does not occur, there
always be a part of the PDF tail for which the amplitudes are too high
for WT to work. Such a ``mild'' breakdown will modify the PDF tail in
a way that may correspond to intermittency.  In fact, this case is
easier to study analytically because WT still works for most of the
PDF and the wave breaking phenomenon can be modeled simply as a
phenomenological cutoff of the PDF tail reflecting the fact that no
waves exist above the breaking amplitude. The wave breaking causes
``leakage'' and, therefore, a flux in the {\em amplitude} space which
is the key phenomenon leading to deviations from the Gaussian
equilibrium and intermittency. Note an analogy with the well-known
{k-space} fluxes (cascades) corresponding to Kolmogorov turbulence
which is qualitatively different from the thermodynamic equilibrium
state.  In this paper we will derive an equation for the wave
amplitude PDF and we will find its steady state solutions
corresponding to the finite flux in the {\em amplitude} space.
Consequently, we will show that the resulting wave fields are
intermittent at each wavenumber with an anomalously large probability
of the large-amplitude waves.

{\bf Definition of RPA fields --- }
  Previously, the random phase approximation (RPA) has typically
assumed that the phases evolve much more rapidly than the amplitudes
and, therefore, there exist time intervals where the phases are random
but the amplitudes are deterministic \cite{ZLF}. However, numerical
simulations indicate that the phase and the amplitude vary at the same
time scale \cite{push}.  Thus, we need to generalize RPA to the case
where both the phases and the amplitudes are random quantities.  Such
generalization was done in \cite{ln} where higher moments for 3-wave
systems were considered. In the present letter, we will be dealing
with 4-wave systems and we will work directly with PDF's rather than
moments.

Let us consider a wavefield $a({\bf x}, t)$ in a periodic box of
volume ${\cal V}$ and let the Fourier transform of this field be $a_k$
where $ k \in {\cal Z}^d$ and $d$ is the space dimension.  Later we
take the large box limit in order to consider homogeneous wave
turbulence. Let us write complex $a_k$ as $a_k =A_k \psi_k $ where
$A_k \in {\cal R}^+ $ is the amplitude and $\psi_k \in {\cal S}^1$ is
a phase factor (${\cal S}^1$ being the unit circle in the complex
plane).  We say the wavefield $a_k$ is of the {\em RPA type} if all
variables in the set $\{ A_k, \psi_k ; k \in {\cal Z}^d \}$ are
statistically independent random variables and $\psi_k$'s are
uniformly distributed on ${\cal S}^1$.
Defined this way RPA refers not only to the phase but also the
amplitude statistics and therefore we suggest a slightly different
reading of this acronym: ``Random Phase and Amplitude''.

The above properties are sufficient for our WT analysis and yet such
fields may be strongly non-Gaussian.  Indeed, RPA allows any shape of
the PDF for amplitudes $A_{k}$ and, therefore, it will be a good tool
for describing intermittency.

{\bf  Weakly nonlinear evolution ---}
Consider a weakly nonlinear wavefield dominated by the 4-wave
interactions, e.g.  the water-surface gravity
waves~\cite{ZLF,hasselman,OsbornePRL,rough}, Langmuir waves in
plasmas~\cite{ZLF,GS} and the waves described by the nonlinear
Schroedinger equation~\cite{DNPZ}.  In the finite box, we have the
following Hamiltonian equations for the Fourier modes of this field,
\begin{equation}
i \dot b_l = \epsilon \sum_{\alpha\mu\nu}
{ W^{l\alpha}_{\mu\nu}}\bar b_\alpha b_\mu b_\nu 
e^{i\omega^{l\alpha}_{\mu\nu}t}
\delta^{l\alpha}_{\mu\nu}
\label{FourWaveEquationOfMotionB}
\end{equation}
where $b_l$ is the wave action variable in the interaction
 representation, $l \in {\cal Z}^d$, $W^{l \alpha} _{\mu \nu} \sim 1$
 is an interaction coefficient,
$\omega^{l\alpha}_{\mu\nu} =
\omega_l+\omega_{\alpha}-\omega_{\mu}-\omega_{\nu}$,
$\omega_l$ is the frequency of mode $l$
and $\epsilon \ll 1$ is a nonlinearity parameter.  We are going expand
in $\epsilon$ and consider the long-time behavior of a wave field, but
in order to make such an analysis consistent we have to renormalize
the frequency of (\ref{FourWaveEquationOfMotionB}) as
\begin{equation}
i \dot a_l = \epsilon \sum_{\alpha\mu\nu}
{ W^{l\alpha}_{\mu\nu}}\bar a_\alpha a_\mu a_\nu 
e^{i\tilde\omega^{l\alpha}_{\mu\nu}t}
\delta^{l\alpha}_{\mu\nu} - \Omega_l a_l,
\label{FourWaveEquationOfMotionA}
\end{equation}
where $a_l = b_l e^{i \Omega_l t} $, $\Omega_l = 2 \epsilon \sum_\mu
 W^{l \mu}_{l \mu} |A_\mu(0)|^2 $ is the nonlinear frequency shift
 arising from self-interactions and 
$\tilde \omega^{l\alpha}_{\mu\nu} = \omega^{l\alpha}_{\mu\nu}
 +\Omega_l+\Omega_\alpha-\Omega_\mu-\Omega_\nu$.

For small nonlinearity, the linear time-scale $2 \pi / \omega$ is a
lot less than the nonlinear evolution time which (as will be evident
below, see e.g. (\ref{pdeofmoment})) is $2 \pi \epsilon^2/ \omega$.
Thus, to filter out fast oscillations at the wave period, let us seek
for the solution at an intermediate time $T$ such that $2 \pi / \omega
\ll T \ll 1/\omega \epsilon^2$.  Now let us use a perturbation
expansion in small $\epsilon$,
$
a_l(T)=a_l^{(0)}+\epsilon a_l^{(1)}+\epsilon^2 a_l^{(2)}.
$
Substituting this in (\ref{FourWaveEquationOfMotionA}) we get in the
zeroth order a time independent result,
$ a_l^{(0)}(T)=a_l(0)\label{definitionofa} $.
For simplicity, we will write $a_l(0)= a_l$, understanding that a
quantity is taken at $T=0$ if its time argument is not mentioned
explicitly.  The first iteration of (\ref{FourWaveEquationOfMotionA})
gives
\begin{eqnarray}
a_l\one(T) = - i \sum_{\alpha\mu\nu} W^{l\alpha}_{\mu\nu} \bar
 a_\alpha a_\mu a_\nu \delta^{l\alpha}_{\mu\nu} \Delta^{l
 \alpha}_{\mu\nu} + i {\Omega_l \over \epsilon} a_l T.  \label{FirstIterate}
\end{eqnarray}
where $\Delta^{l \alpha}_{\mu\nu} \equiv \Delta^{l \alpha}_{\mu\nu}(T) =
({e^{i\tilde\omega^{l\alpha}_{\mu\nu}T}-1})/{i \tilde\omega^{l\alpha}_{\mu\nu}}.
 \label{NewellsDelta}$
Iterating one more time we get
\end{multicols}
\leftline{---------------------------------------------------------------------
----}
\begin{eqnarray}
a_l\two(T)=\sum_{\alpha\beta\mu\nu v u}\left( 
W^{\mu\nu}_{\alpha u} W^{lu}_{v\beta} \delta^{\mu\nu}_{\alpha u} \delta^{lu}_{v\beta}
 a_{\alpha} a_v a_\beta 
\bar a_{\mu} \bar a_{\nu} 
E(\tilde\omega^{l\mu\nu}_{\alpha v \beta}, \tilde\omega^{lu}_{v\beta}) 
- 
2 W^{\alpha v}_{\mu\nu} W^{l u}_{v \beta}
\delta^{\alpha v}_{\mu\nu} \delta^{l u}_{v \beta}
\bar a_{\alpha} \bar a_{u}
a_{\mu} a_{\nu} a_\beta 
E(\tilde \omega^{l\alpha u}_{\mu\nu\beta}, \tilde \omega^{lu}_{v\beta})\right) 
-\Omega_l^2 a_l \frac{T^2}{2\epsilon^2}
\nonumber  \\ 
+ \frac{1}{\epsilon}
\sum_{\alpha\mu\nu}\left(
\Omega_l 
W^{l\alpha}_{\mu\nu}
\delta^{l\alpha}_{\mu\nu}
\bar a_{\alpha} a_{\mu} a_{\nu} 
E(\tilde\omega^{l\alpha}_{\mu\nu},0)  
-W^{l \alpha}_{\mu\nu }
\delta^{l \alpha}_{\mu\nu }\bar{ a_{\alpha}} 
a_{\mu} a_{\nu}(\Omega_\alpha
-2\Omega_\nu) 
\int \limits_0^T
\tau e^{i\tilde\omega^{l \alpha}_{\mu\nu } \tau} d \tau
\right)
, \hbox{\ with\ }
E(x,y)=\int_0^T \Delta(x-y)e^{i y t} d t .
\label{SecondIterate}
\end{eqnarray}
\rightline{---------------------------------------------------------------------
----}
\begin{multicols}{2}

{\bf Evolution of statistics --- } We will now develop a statistical
description via averaging over the initial fields $a_k(0)$ which are
taken to be of the RPA type.  Of course, to have a non-trivial
description valid over the nonlinear evolution time, the fields must
remain of the RPA type over the nonlinear time in the leading order in
$\epsilon$. The proof of this involves considering the full
multi-particle PDF and will be published separately because it is
rather lengthy and outside of the scope of the present paper
\cite{physd}.  Let us introduce a generating function $ Z(\lambda
,t)=\langle e^{\lambda |a_k(t)|^2}\rangle, $
where $\lambda$ is a real parameter.  Then PDF of the wave intensities
$s= |a_k(t)|^2$ at each ${\bf k}$ can be written as an inverse Laplace
transform,
$
P(s,t) = 
\langle \delta (|a_k(t)|^2 -s) \rangle =
{1 \over 2 \pi i}  \int_{-i \infty}^{+i \infty} Z(\lambda, t)e^{-s
\lambda}d\lambda. \label{distribution}
$
For the one-point moments  we have
\begin{eqnarray} 
M_k^{(p)} \equiv \langle
|a_k|^{2p}\rangle 
=\langle
|a|^{2p}e^{\lambda |a|^2}\rangle|_{\lambda=0}=\nonumber\\Z_{\lambda \cdots \lambda}|_{\lambda=0}
= \int_{0}^{\infty} s^p P(s, t) \, ds \label{momPDF}, 
\end{eqnarray}
where $p \in {\cal N}$ and subscript $\lambda$ means differentiation
with respect to $\lambda$ $p$ times.

At $t=T$ we have 
\begin{eqnarray}
Z(T) = \langle e^{\lambda|a_k^{(0)} +\epsilon a_k^{(1)} +\epsilon^2
a_k^{(2)}|^2}\rangle  = \hspace{1.5cm} \nonumber \\
\langle e^{\lambda
|a_k^{(0)}|^2} \langle 1+\lambda \epsilon (a_k^{(1)} \bar a_k^{(0)}+
{\rm cc})
\nonumber \\
 \lambda
\epsilon^{2}(|a_k^{(1)}|^{2} + (a_k^{(2)}\bar a_k^{(0)} +{\rm cc}) 
) + 
\nonumber \\
{\lambda^2 \e^2 \over 2} (a_k^{(1)}\bar a_k^{(0)} +{\rm cc}
)^2
\rangle_{\psi} \rangle_{A}
\nonumber \\
=Z(0) + \e \lambda \langle e^{\lambda
|a_k^{(0)}|^2} \langle a_k^{(1)} \bar a_k^{(0)}+
{\rm cc} \rangle_{\psi} \rangle_{A} + \nonumber \\
\epsilon^{2} \langle \langle (\lambda + \lambda^2 A^2) |a_k^{(1)}|^{2} 
+ \lambda (a_k^{(2)}\bar a_k^{(0)} +{\rm cc} ) \nonumber \\ 
+ {\lambda^2  \over 2} (a_k^{(1)2} \bar a_k^{(0)2} +{\rm cc})
\rangle_{\psi} \rangle_{A} \hspace{1.5cm}
\label{zmoment} 
\end{eqnarray}
where ${\rm cc}$ stands for complex conjugate of the previous terms
and $\langle\dots\rangle_\psi$ and $\langle\dots\rangle_A$ denote
phase and amplitude averaging respectively. Note that in RPA fields
the phases and the amplitudes are statistically independent so that
these two averaging could be done independently. First let us
substitute $a_k^{(1)}$ and $a_k^{(2)}$ from (\ref{FirstIterate}) and
(\ref{SecondIterate}) respectively and perform the phase
averaging. For the terms proportional to $\epsilon$ we have
\begin{eqnarray}
\langle a_k\one\bar a_k^{(0)}\rangle_\psi= \nonumber \\
 - i \sum_{\alpha\mu\nu} W^{k\alpha}_{\mu\nu} 
\langle \bar a_k \bar
 a_\alpha a_\mu a_\nu\rangle_\psi
 \delta^{k\alpha}_{\mu\nu} \Delta^{k
 \alpha}_{\mu\nu} + i \Omega_l \langle |a_k|^2 \rangle  T\nonumber\\
 =- 2 i \sum_{\alpha} W^{k\alpha}_{k\alpha} 
 A_k^2 A_\alpha^2 \cdot T
+ i \Omega_l A_k^2 T.
\label{EpsilonTerms}
\end{eqnarray}
where we have used the fact that $\Delta(0)=T$ and we have used the
RPA's ``Wick's Theorem''
$$\langle \bar a_k \bar  a_\alpha a_\mu a_\nu\rangle_\psi =
A_k^2 A_\alpha^2 (
\delta^k_\mu \delta^\alpha_\nu +\delta^k_\nu\delta^\alpha_\mu).
$$ 
Note that the above expression should also contain the {\it singular
cumulant}, i.e. term $(A_\alpha^4 - 2A_\alpha^2) \delta^\alpha_\nu
\delta^\alpha_\mu \delta^\alpha_k$, see \cite{ln}. However we do not
write this term here, since its contribution is of the order of $1/N$
smaller because it has one extra delta function.

We see from (\ref{EpsilonTerms}) that the choice 
\begin{equation}
\Omega_k  =2 \sum_{\alpha} W^{k\alpha}_{k\alpha} A_\alpha^2
\label{FrequencyRenormalization}
\end{equation}
makes the contribution of $\langle a_k\one\bar a_k^{(0)}\rangle_\psi$ terms 
to be equal to zero. 

We therefore obtain
\begin{eqnarray}
Z(T) -Z(0) = \hspace{2.5cm} \nonumber\\ 
\lambda
\epsilon^{2}
\Big\langle
\big\langle
\lambda |a_k^{(1)}|^{2} (1+\lambda |a_k|^2)+
a_k^{(2)}\bar a_k^{(0)} + 
\bar a_k^{(2)} a_k^{(0)}+\nonumber\\ 
+\frac{\lambda}{2} ( (\bar a_k^{(0)} a_k\one)^2+{\rm cc}) 
\big\rangle_{\psi} \Big\rangle_{A}. \label{Zintermediate}
\end{eqnarray}
where as an
intermediate result we have
\begin{eqnarray}
\langle \bar a_k^{(0)} a_k\two\rangle_\psi=\hspace{2.5cm}
\nonumber\\
2\sum\limits_{\alpha \mu\nu}
\delta^{k\alpha}_{\mu\nu}
|W^{k\alpha}_{\mu\nu}|^2 A_k^2 
( A_\mu^2 A_\nu^2 - 2 A_\alpha^2 A_\mu^2)
 E(0,\tilde\omega^{k\alpha}_{\mu\nu})\nonumber\end{eqnarray}
and
\begin{eqnarray}
\langle |\bar a_k\one|^2\rangle_\psi=
\sum\limits_{\alpha \mu\nu}\delta^{k\alpha}_{\mu\nu}
|W ^{k\alpha}_{\mu\nu}|^2 A_\alpha^2 A_\mu^2 A_\nu^2 
|\Delta(\tilde\omega^{k\alpha}_{\mu\nu})|^2.
\nonumber
\end{eqnarray}
Here terms proportional to $T^2$ drop from $\langle \bar a_k^{(0)}
a_k\two\rangle_\psi$ and $\langle |\bar a_k\one|^2\rangle$ because of
the choice (\ref{FrequencyRenormalization}) of frequency
renormalization.  
Furthermore, 
\begin{eqnarray}
\langle
(a_k\one\bar a_k^{(0)})^2                                                              
\rangle_\psi=\nonumber\\
-2 W^{k k }_{k k } A_k^6 T^2 \Omega_k
-A_k^4 \Omega_k^2 T^2 - 4 \sum_\alpha(W^{k\alpha}_{k\alpha}
A_k A_\alpha T )^2\nonumber
\end{eqnarray}

To complete the derivation of the equation for the time evolution of
the generating function $Z(T)$ we have to take a large box limit,
which implies that sums will be replaced with integrals, the Kronecker
deltas will be replaced with Dirac's deltas, $\delta^{l \alpha}_{m
n}\to\delta^{\alpha l}_{mn}/{\cal V}$, where we introduced short-hand
notation, $\delta^{\alpha l}_{mn}=\delta(k_\alpha+k_l-k_m-k_n)$.  Then
(\ref{Zintermediate}) will still hold, but with
\begin{eqnarray}
\langle \bar a_k^{(0)} a_k\two\rangle_\psi= \hspace{2.5cm}
\nonumber\\
2\int d 123
\delta^{k1}_{23}
|W^{k1}_{23}|^2 A_k^2 
( A_2^2 A_3^2 - 2 A_1^2 A_2^2)
 E(0,\tilde\omega^{k1}_{23})\nonumber\end{eqnarray}
and
\begin{eqnarray}
\langle |\bar a_k\one|^2\rangle_\psi=
\int d 123\delta^{k1}_{23}
|W ^{k1}_{23}|^2 A_1 A_2 A_3 
|\Delta(\tilde\omega^{k1}_{23})|^2.
\nonumber
\end{eqnarray}
We also have
\begin{eqnarray}
\langle
(a_k\one\bar a_k^{(0)})^2
\rangle_\psi = 0, 
\end{eqnarray}
because this terms will be $1/N$ times smaller than $\langle |\bar
a_k\one|^2\rangle_\psi$ and $\langle \bar a_k a_k\two\rangle_\psi$
terms because it has one less summation index. Therefore it vanishes
in the $N\to\infty$ limit.

Further we take a large $T$ limit, and take into account that
$$\lim\limits_{T\to\infty}E(0,x)= (\pi
\delta(x)+iP(\frac{1}{x})) T,$$
and
$$\lim\limits_{T\to\infty}|\Delta(x)|^2=2\pi T\delta(x),$$
(see e.g. \cite{Ben}).

Finally we perform amplitude averaging, noticing that 
$$Z(0)=\langle e^{\lambda |a_k|^2}\rangle_A,$$
and
$$\frac{\partial Z}{\partial \lambda} = \langle |a_k|^2 e^{\lambda |a_k|^2}
\rangle_A.$$
to obtain
\begin{equation}
Z(T) = Z(0) + \epsilon^2T\cdot( 
\lambda \eta Z +(\lambda^2 \eta - \lambda \gamma)
 Z_\lambda).\label{pdeofmoment2}
\end{equation}
Approximating $(Z(T)-Z(0))/T$ by $\dot Z$, we have
\begin{equation}
\dot Z = \lambda \eta Z +(\lambda^2 \eta - \lambda n \gamma)
 Z_\lambda,\label{pdeofmoment}
\end{equation}
where
\begin{eqnarray}
\eta_k &=& 4 \pi \epsilon^2 \int
|W^{k1}_{23}|^2 \delta^{k1}_{23}
\delta(\omega^{k1}_{23}) n_1 n_2 n_3 \, d123,\nonumber
\\ 
\gamma_k &=&
8 \pi \epsilon^2 \int  |W^{k1}_{23}|^2 \delta^{k1}_{23}
\delta(\omega^{k3}_{12}) 
 \Big[ n_1 (n_2 + n_3) - n_2 n_3\Big] \, d123,\nonumber
\end{eqnarray}
here wavenumbers $k, k_1, k_2, k_3 \in {\cal R}^d$, $\delta$'s now
mean Dirac $\delta$-functions, $n_{1,2,3} \equiv n(k_{1,2,3}) $ and
$d123 = d {\bf k_1} d {\bf k_2} d {\bf k_3},$ Differentiating
(\ref{pdeofmoment}) with respect to $\lambda$ $p$ times we get the
evolution equation for the moments:
$$ \dot M^{(p)}_k = -p \gamma_k
M^{(p)}_k + p^2 \eta_k M^{(p-1)}_k,\label{MainResultOne} $$
which, for $p=1$ gives the standard kinetic equation, $ \dot n_k = -
\gamma_k n_k + \eta_k . $ First-order PDE (\ref{pdeofmoment}) can be
easily solved by the method of characteristics.  Its steady state
solution is $$ Z=(1 -\lambda n_k)^{-1} $$ which corresponds to the
Gaussian values of momenta $M^{(p)} = p! n_k^p$.  However, these
solutions are invalid at small $\lambda$ and high $p$'s because large
amplitudes $s=|a|^2$, for which nonlinearity is not weak, strongly
contribute in these cases. Due to the integral nature of definitions
of $M^{(p)}$ and $Z$ with respect to the $s=|a|^2$, the ranges of
amplitudes where WT is applicable are mixed with, and contaminated by,
the regions where WT fails. Thus, to clearly separate these regions it
is better to work with quantities which are local in $s=|a|^2$, in
particular the probability distribution $P(s)$.  Taking the inverse
Laplace transform of (\ref{pdeofmoment}) we have
\begin{equation}
\dot P+\partial_{s}F=0, \label{peqn}
\end{equation}
 where $ F=-s(\gamma P+\eta\partial_{s}P) $ is a probability flux in
the s-space.
Consider the steady state solutions, $\dot P =0$, 
\begin{equation}
-s(\gamma P+\eta\partial_{s}P) = F= \hbox{const}. \label{f} 
\end{equation}
Note that in the steady state $\gamma /\eta = n_k$ which follows from
kinetic equation.  The general solution to (\ref{f}) is
$$P=P_{hom} + P_{part}$$
 where 
$$P_{hom} = \hbox{const} \, \exp{(-s/n)}$$ 
is the general solution to the homogeneous equation (corresponding to
$F=0$) and $P_{part}$ is a particular solution, 
$$P_{part} = -({F}/{\eta}) Ei({s}/{n}) \exp{(-s/n)} $$
where $Ei(x)$ is the integral exponential function.

At the tail of the PDF, $s \gg n_k$, the solution can be represented
as series in $1/s$,
$
P_{part} = - F/( s\gamma)-\eta F/(\gamma s)^2+\cdots.
$
Thus, the leading order asymptotic of the finite-flux solution is
$1/s$ which describes strong intermittency.

Note that if the weakly nonlinearity assumption was valid uniformly to
$s=\infty$ then we had to put $F=0$ to ensure positivity of $P$ and
the convergence of its normalization, $\int P \, ds =1$. In this case
$P=P_{hom} = n \, \exp{(-s/n)}$ which is a pure Rayleigh distribution
corresponding to the Gaussian wave field.  However, WT approach fails
for the amplitudes $s \ge s_{nl}$ for which the nonlinear time is of
the same order or less than the linear wave period and, therefore, we
can expect a cut-off of $P(s)$ at $s = s_{nl}$. An estimate based on
the dynamical equation (\ref{FourWaveEquationOfMotionB})
gives\footnote{This estimate assumes that if the wave amplitude at
some $k$ happened to be of the critical value $s_{nl}$ then it will
also be of similar value for a range of $k$'s of width $k$. In other
words, strong nonlinearity widens the $k$-space correlation from zero
(RPA value) to $k$ (value for the coherent structures involved in the
wave breaking).}  $s_{nl}= \omega/\epsilon W k^2 $.  This
phenomenological cutoff can be viewed as a wave breaking process which
does not allow wave amplitudes to exceed their critical value, $P(s)
=0$ for $s > s_{nl}$.  Now the normalization condition can be
satisfied for the finite-flux solutions.  However, having a constant
negative flux $F<0$ corresponds to a source at $s=s_{nl}$ which
dictates the necessity of a sink for some $s < s_{nl}$ to preserve the
normalization of $P(s)$. Note however that the probability sink does
not have to correspond to any physical ``removal'' of waves with
certain amplitudes. The sink should be present solely because the
probability is diluted due to acceptance of new members with
$s=s_{nl}$ into the statistical ensemble. In this case, the sink must
be proportionate to the probability and, taking into account the
normalization condition, we can write a modified equation for the PDF
in the presence of cutoff,
\begin{equation}
\dot P- \partial_{s} (s\gamma P+s\eta\partial_{s}P)= - F_*, \label{peqnM}
\end{equation}
with $F_* = - P(s_{nl}) \gamma /s_{nl} $.
The general solution solution to this equation is
$
 P = [C  -{{F_*} } Ei({s}/{n} - \log s)/\eta ] \exp{(-s/n)},
$
there constant $C$ is fixed by the normalization condition.  This
solution is close to the Rayleigh distribution in the PDF core, $s
\sim n$, and it has a $1/s$ tail at $n \ll s < s_{nl} $.

{\bf Discussion ---}
We found that the WT intermittency shows as an anomalously high ($\sim
1/s$) probability of the large-amplitude waves whereas at lower
amplitudes distribution appears to be close to Rayleigh ($\sim
e^{-s/n}$) which corresponds to Gaussian wave fields.  We showed that
wave breaking is essential for WT intermittency to be present in the
system, yet the details of wave breaking are not important. The role
of wave breaking is just to ensure that no wave can have amplitude
greater than critical value $s_{nl}$. This simple condition leads to
huge mathematical consequences as it generates the flux solutions in
the amplitude space and therefore creates the $ 1/s$ intermittency.
On the other hand, the amplitude of the $1/s$ tail is not prescribed
by WT and will depend on a particular wave breaking mechanisms in a
particular system.  However, some conclusions about the dependence of
the tail amplitude on the physical parameters can be reached using a
dimensional arguments.

Consider a classical example of the gravity waves on surface of deep
water.  The linear dispersion relation is given by $\omega_k = \sqrt{
g k}$, and the coefficient of nonlinear interaction
$W^{k{\bf \alpha}}_{\bf \mu\nu}$
is given in \cite{ZLF}. This system has two power-law steady state
solutions.  First one is the spectrum corresponding to the direct
cascade of energy toward high-wave numbers, $n_k \propto k^{-4}$
\cite{ZLF,Zakfil}. Second one is the spectrum corresponding to the
inverse cascade of wave action toward the small $k$ values,
$n_k\propto k^{-23/6}$.  In addition to the gravity constant $g$, the
only quantity which determines the state of the system in the direct
cascade range is the energy flux ${\cal P}$ whereas in the inverse
cascade range - the particle flux $\cal Q$. The PDF tail strength can
be characterized by its area which is a dimensionless number and,
therefore, has to depend on the relevant dimensionless combinations in
the direct and the inverse cascade ranges, ${\cal P}^{2/3} k^{1/3} /g$
and ${\cal Q} k/g$ respectively. Thus, the PDF tail thickness grows
with $k$ but its length descreases until it completely disappears at
$k \sim k_{nl}$ (equal to $g^3/{\cal P}^2$ and $g/{\cal Q}$
respectively).

This effect is illustrated in Figure 1 which shows PDF's obtained by
numerical simulations of the surface waves on deep water forced at low
$k$'s and dissipated at high $k$'s. Pseudospectral numerical method
similar to that of \cite{yoko},\cite{OsbornePRL} was used on a 256x256
grid.

At moderate wavenumber ($k=15 k_{min}$) one can see a PDF tail in the
range $4 n_k < s < 10 n_k$ characterized by an order of magnitude
enhanced probabilities with respect to the Rayleigh distribution.
Unfortunately the range of $s$ where PDF converged to a stable value
in this experiment was not large enough to reach $s \gg n$ values and,
therefore, for an asymptotic scaling to develop.  To increase this
range a much longer computing to gain good statistics of very rare
events at the PDF tail is necessary, which we can not perform with our
resources.
 
At a higher wavenumber ($k=35 k_{min}$) one can see that the large
amplitude waves are less probable than the ones predicted by the
Rayleigh distribution.  This is because the wavebreaking happens now
closer to the PDF core causing the PDF cut-off seen at the figure.

In this letter we considered WT which is weak on average so that the
wave breaking occurs only in the PDF tail, i.e. $s_{nl} \gg n$.  It
does not apply to the cases when, at some large $k$, the wave breaking
may become so strong that it occurs for most of the waves in the PDF
core.  These cases where predicted and discussed in \cite{biven}, but
their statistics would be hard to describe analytically because of the
strong nonlinearity.

{\bf Acknowledgments ---} YL is supported by NSF CAREER grant DMS
0134955 and by ONR YIP grant N000140210528

\begin{figure}
\epsfxsize=8cm\epsffile{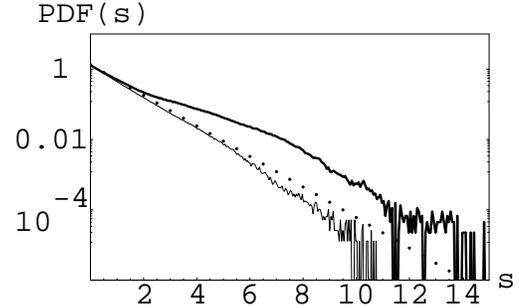}
\caption[]{\footnotesize\baselineskip11pt PDF of $|a_k|^2$ for $k=15
 k_{min}$ (thick curve) and $k=35 k_{min}$ (thin curve)
 and their comparison with Rayleigh distribution (dotted
 line). Amplitude s is normalised so that the two curves have the same
 slope at $s=0$.}
\end{figure}

\end{multicols}

\end{document}